\documentclass[prl,twocolumn,showpacs]{revtex4}
\usepackage{graphicx}
\usepackage{dcolumn}
\usepackage{amssymb}
\usepackage{bm} 
\def\Q{\mbox{\sffamily\bfseries Q}}

\def\I{\mbox{\sffamily\bfseries I}}
\def\A{\mbox{\sffamily\bfseries A}}

\def\G{\mbox{\sffamily\bfseries G}}

\def\f{\mbox{\sffamily\bfseries f}}

\def\nabbold{\mbox{\boldmath $\nabla$\unboldmath}}

\def\sigbold{\mbox{\boldmath $\sigma$\unboldmath}}

\def\sigac{\mbox{\boldmath $\sigma$\unboldmath$^a$}}

\def\sigv{\mbox{\boldmath $\sigma$\unboldmath$^v$}}
\def\sigop{\mbox{\boldmath $\sigma$\unboldmath$^{\tiny OP}$}}

\def\beq{\begin{equation}}                           
\def\eeq{\end{equation}}                           
\def\bea{\begin{eqnarray}}                           
\def\eea{\end{eqnarray}}                           
\begin{document}

\textwidth = 6.5 in
\textheight = 9 in
\oddsidemargin = 0.0 in
\evensidemargin = 0.0 in
\topmargin = 0.0 in
\headheight = 0.0 in
\headsep = 0.0 in
\parskip = 0.2in
\parindent = 0.0in

\title{Rheology of Active-Particle Suspensions}
\author{Yashodhan Hatwalne$^{1}$, Sriram Ramaswamy$^{2}$, Madan Rao$^{1,3}$ 
and R. Aditi Simha{$^{2,4}$}}
\affiliation{$^{1}$Raman Research Institute, C.V. Raman Avenue, Bangalore 560 080 
INDIA\\ 
$^2$Centre for Condensed Matter Theory, Department of Physics,
Indian Institute of Science, Bangalore 560 012 INDIA\\
$^3$ National Centre for Biological Sciences, UAS-GKVK Campus, Bellary Road,
Bangalore 560 065 INDIA\\
$^4$MPI-PKS, N\"{o}thnitzer Str. 38, 01187 Dresden, Germany}
\date{version of 25 Aug 2003, printed \today}
\begin{abstract}We study the interplay of activity, order and flow 
through a set of coarse-grained equations governing the 
hydrodynamic velocity, concentration and stress fields in a suspension of active, 
energy-dissipating particles. We make several predictions for the rheology 
of such systems, which can be tested on bacterial suspensions, cell extracts 
with motors and filaments, or artificial machines in a fluid. 
The phenomena of cytoplasmic streaming, elastotaxis and active mechanosensing find 
natural explanations within our model. 
\end{abstract} 
\maketitle

An active particle \cite{active,tonertu} absorbs energy from its surroundings or from 
an internal fuel tank and dissipates it in the process of carrying out internal 
movements usually resulting in translatory or rotary 
motion. This broad definition includes macroscopic machines and organisms, 
living cells, and their components such as actin-myosin and 
ion pumps \cite{alberts}. 
In this paper, we consider the interplay of activity, order and flow 
via coarse-grained equations governing the 
hydrodynamic velocity, concentration and stress fields in a suspension containing active 
particles of linear size $\ell$, at concentration $\phi$, each particle exerting a typical 
force $f$ on the ambient fluid, with the activity of an individual particle 
correlated over a time $\tau_0$ (say the `run' time of a bacterium), and collective 
fluctuations in the activity correlated over length scales $\xi$ and timescales $\tau$ . 
Rather than focussing on ordered phases 
\cite{sppwork}, instabilities \cite{sppwork,tanniecris}, or 
patterns (asters, vortices, 
spirals) formed in such assemblies \cite{asters} 
which our equations are of course capable of predicting, we apply them 
in the isotropic phase, with a view to understanding  
how a system such as a biological cell, 
composed of {\em active} elements, responds to deformation or mechanical stress.  
In addition to 
throwing light on full-cell rheometry \cite{crick,fabry}, our equations form the 
framework for an analysis of any experiment probing the mechanical consequences 
of biological activity.  

Our simple model makes rather interesting predictions: 
An orientationally ordered state
of active particles has a nonzero,  
macroscopic, anisotropic stress in contrast with thermal equilibrium nematics. 
Activity contributes an amount 
$\delta \eta \sim f \ell c_0 \tau$ to the viscosity, with a sign determined 
by the type of active particle, and always enhances  
the apparent (noise) temperature.  
The latter greatly enhances the amplitude of the 
$t^{-d/2}$ long-time tails \cite{tails} in the  
velocity autocorrelation.  
On approaching an {\em orientationally} ordered state, 
active suspensions with $\delta \eta >0$ behave like passive systems near 
{\em translational} freezing, showing strong shear thickening 
and Maxwell-like viscoelasticity. 
Nonlinear fluctuation corrections give  
a dynamic modulus $G^*(\omega) \sim \sqrt{i\omega}$ for $\omega \gg \tau^{-1}$
observable over a large dynamic range, since $\tau$ is large. 
Cytoplasmic streaming \cite{bray}, in which material flows from the 
depolymerising trailing edge to the polymerising leading edge of a crawling 
amoeboid cell, finds a natural explanation in our model, as do 
elastotaxis \cite{fontes} and active mechanosensing \cite{mechanosens}, 
where cells orient their motion along preferred axes of the ambient medium. 

These results follow from equations of motion based simply on the conservation law    
$\partial_t {\bf g} = - \nabbold \cdot \sigbold$  
for the total (particles $+$ fluid) momentum density 
${\bf g}({\bf x},t)$ for an incompressible suspension.    
The stress tensor $\sigbold$ must in turn be determined 
by constitutive relations which emerge 
from an additional equation of motion 
for an active order parameter field. We ignore, for simplicity, 
the dynamics of the active-particle concentration $\phi({\bf x},t)$, 
energy density and nutrient fields.  

To determine the contributions of activity to the stress, we need the 
forces associated with the active particles.  
Since there are no {\em external} forces on the system, 
the simplest active particle, on long timescales, is a permanent force dipole 
\cite{brennen} 
\begin{figure}
\includegraphics[width=6cm,height=4cm]{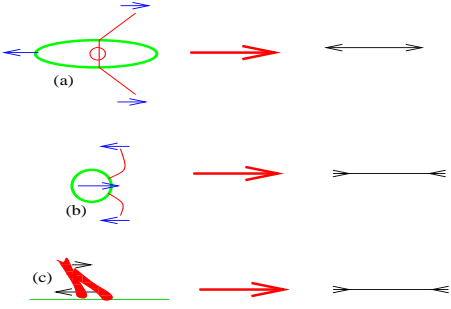}
\caption{\label{dipoles} Force dipoles for (a) a rowboat, (b) a ``bacterium'' 
with two flagella \cite{prnott}, and (c) a motor on a filament }
\end{figure}
(see Fig. \ref{dipoles}).  
A collection of such particles \cite{sppwork} 
where the $\alpha$th particle,  
centred at 
${\bf R}_{\alpha}$, 
has point forces of strength $f$ and directions $\pm \hat{\bf n}_{\alpha}$ situated 
at ${\bf R}_{\alpha}+ b \hat{\bf n}_{\alpha}$ and
${\bf R}_{\alpha}+ b' \hat{\bf n}_{\alpha}$, leads to a force density  
\bea
\label{forceleading}
{\bf F}^a &\simeq& 
-(b + b') f
\nabbold\cdot \sum_{\alpha} \hat{\bf n}_{\alpha}\hat{\bf n}_{\alpha}
\delta({\bf r} - {\bf R}_{\alpha}) \nonumber \\
&+&{(b + b')(b - b') \over 2} f
\nabbold \nabbold : \sum_{\alpha} \hat{\bf n}_{\alpha}\hat{\bf n}_{\alpha} 
\hat{\bf n}_{\alpha} \delta({\bf r} - {\bf R}_{\alpha}) + \ldots \nonumber \\
&=& \nabbold \cdot \sigac,    
\eea
which defines the active stress $\sigac$.  
For bacteria swimming at speed $v_0$ in a fluid of viscosity $\eta_0$, $f \sim \eta_0 b v_0$.  
Both polar ($b \neq b'$) and apolar ($b = b'$) particles disturb the fluid;  
the former (a ``mover'') induces a nonzero fluid velocity at its centre and hence moves, 
the latter (a ``shaker''), by symmetry, cannot. 
Note that $\sigac$ is insensitive, at lowest order in gradients, to 
the asymmetry $b-b'$: movers and shakers have the same {\em far-field} 
fluid flow, and an isotropic collection of either or both should 
have similar rheology. Since the force dipole determines an axis for each active 
particle, the natural definition $\phi({\bf r})\Q({\bf r}) \equiv  
\sum_{\alpha} (\hat{\bf n}_{\alpha}\hat{\bf n}_{\alpha} - {1 \over 3} \I)
\delta({\bf r} - {\bf R}_{\alpha})$   
(where $\I$ is the unit tensor) of a local nematic order 
parameter or alignment tensor 
$\Q$ associated with the activity lets us explore the rheological 
consequences of spatiotemporal correlations in the activity 
by a simple generalisation of nematodynamics \cite{degp}.  
%
We have thus established that the active contribution to the deviatoric 
(traceless symmetric)  
stress \cite{QQ}
\beq
\label{sigproptoq}
\sigac - (1/3)\I Tr \sigac =W \Q + W_2\Q^2 + .... 
\eeq 
where the constants $W, \, W_2 \sim (b + b') f \phi$ characterise the strength 
of the elementary force dipoles, and the sign of $W$ has vital rheological consequences.  
The relation (\ref{sigproptoq}) is at the heart of the novel mechanical properties 
of active systems \cite{sppwork}.  
Even without equations of motion, (\ref{sigproptoq}) tells us why an active 
suspension with long-range nematic order is different from its passive 
counterpart. Both have $\Q\neq 0$; the passive nematic, bound by 
Pascal's Law since it is an equilibrium liquid despite its orientational order, 
has a purely {\em isotropic} mean stress, i.e., a pressure, whereas the active nematic 
has a nonzero mean deviatoric stress, a truly nonequilibrium 
effect.   

\begin{figure}
\includegraphics[width=6cm,height=4cm]{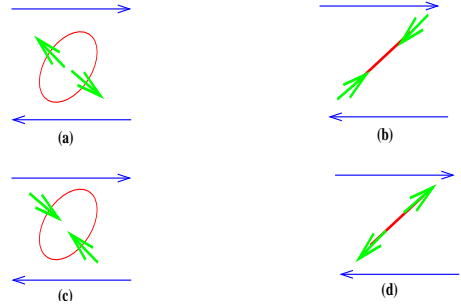}
\caption{\label{stabdestab} Discs (a) and (c) and rods (b) and (d) with active 
force densities attached along their symmetry axes, under shear (horizontal arrows). 
The parameter $W >0$ in (a) and (b) 
and $<0$ in (c) and (d).} 
\end{figure}
Fig. \ref{stabdestab} shows what the parameter $W$ means.  
In an imposed flow, in 
the {\em absence} of activity, discs (rods) tend to spend most of their time with 
symmetry axis along the compression (extension) axis of the flow \cite{forsterprl}. 
When activity is switched on, the flow induced by the intrinsic force dipoles 
will clearly oppose the imposed flow in cases (a) and (b), and enhance it 
in (c) and (d). 

For passive nematogens, $\Q$ is governed by a free-energy functional 
$F[\Q]$ containing polynomials in $\Q$ as 
well as Frank elastic terms $\sim \nabla \Q \nabla \Q$, giving rise to a passive 
order-parameter stress 
\cite{forsterprl} 
\beq
\label{passiveopstress}
\sigop = 
 3 \G - \G \cdot \Q - \Q \cdot \G
\eeq 
where $\G\equiv -\delta F / \delta \Q + (1/3) \I Tr \delta F / \delta \Q$ 
is the nematic molecular field. 
The {\em mean} deviatoric 
passive stress (\ref{passiveopstress}) is zero in both isotropic and 
nematic phases. For small nematic perturbations $\delta \Q$ in the isotropic phase, 
$F \propto a \int  \phi Tr (\delta \Q)^2$  
so that the stress fluctuation $\sim a \phi \delta \Q$ with a coefficient $a$ which 
decreases on approaching the transition to the ordered phase. 
In {\em active} systems, the relation (\ref{sigproptoq}) between stress and 
order parameter does not arise from a free-energy functional, and 
the proportionality constant $W$ has no reason to decrease with increasing nematic 
correlations.
This difference will be seen to be crucial when we compare
the pretransitional viscoelasticity of passive and active nematogenic suspensions.  

Including the viscous stress $\sigv = -\eta_0 \A +  
O(\Q \nabbold {\bf u})$, expressed in terms  
of the rate of deformation $\A \equiv (1/2)[\nabbold {\bf u} + (\nabbold {\bf u})^T]$   
and the hydrodynamic velocity field ${\bf u} \equiv {\bf g}/\rho$ 
for a system of density $\rho$, 
the total deviatoric stress $\sigbold$ in the active case can 
be written as 
$\sigbold = \sigac + \sigv + \sigop$, 
plus a noise source unconstrained 
by a fluctuation-dissipation theorem since this is a nonequilibrium system.  
This defines completely the equation of motion 
for the momentum density ${\bf g}$.  
The coarse-grained equation of motion for $\sigbold$ follows from that for $\Q$ 
which when {\em linearised} involves only terms \cite{sppwork} 
of a form present in passive 
nematodynamics \cite{degp}: 
\beq
\label{opeom}
{\partial \Q \over \partial t} 
  =    - {1 \over \tau} \Q + D \nabla^2 \Q 
              + \lambda_0 \A + .... 
+ \f, 
\eeq
where 
$\tau$ is the activity correlation time, 
$D$ is a diffusivity which in passive systems would be the ratio of 
a Frank constant to a viscosity, $\lambda_0$ is a ``reversible'' kinetic coefficient 
\cite{forsterprl},  
$\f$ is a traceless, 
symmetric, spatiotemporally white tensor noise with variance $N_Q$, 
representing thermal or active fluctuations, and the 
ellipsis includes the coupling of orientation to flow.  

We are now ready to calculate the linear viscoelastic properties of our 
active suspension. 
In the isotropic phase, Eqs. (\ref{sigproptoq}), 
(\ref{passiveopstress}) and (\ref{opeom}), linearised and applied to 
spatially uniform oscillatory shear flow at frequency $\omega$ in the $xy$ plane, 
imply  
\bea
\label{stressvsrate}
\sigma_{xy}(\omega) &=& -\left[\eta_0 + {(a + W) \lambda_0 \over -i \omega + \tau^{-1}}\right] A_{xy} \nonumber \\
&\equiv& -{G'(\omega) - i G''(\omega) \over \omega} i A_{xy}.  
\eea
which defines the storage and loss moduli $G'(\omega)$ and $G''(\omega)$. 
This is the claimed active enhancement or reduction 
$\eta_{act} \propto W \tau$ of the effective 
viscosity at zero shear-rate and zero frequency.  
Activity enhances viscosity in Fig. \ref{stabdestab} (a) and (b), since $W>0$, 
and reduces it in (c) and (d) ($W<0$). 
For $W > 0$ (\ref{stressvsrate}) tells us that 
the viscosity grows substantially as the system 
approaches a transition to orientational order (which is in general 
continuous for active {\em vectorial} order \cite{tonertu}), i.e., as $\tau$ 
is increased. By contrast, in a passive system approaching a nematic phase 
the excess viscosity $\sim a \tau$ is roughly constant since $\tau \propto 1/a$.  
 
Eq. (\ref{stressvsrate}) also predicts strong viscoelasticity  
as $\tau$ increases.  
For {\em passive}  systems $W = 0$. Since $a \propto \tau^{-1}$, 
$G'(\omega \tau \gg 1)$ decreases as  
$\lambda_0  \eta_0 / \tau$.  
There is little viscoelasticity near an {\em equilibrium}  
isotropic-nematic transition. For {\em active} systems, by contrast, 
$W$ is independent 
of $\tau$ and of proximity to the transition. Thus, as $\tau$ grows, 
\beq
\label{gpact}
G'(\omega \tau \gg 1) \simeq W
\eeq
independent of $\tau$ and, of course, the dynamic range over 
which elastic behaviour is seen increases.  
At {\em equilibrium}, one would expect such strong viscoelastic behaviour 
from a fluid or suspension near {\em translational} freezing, not 
near {\em orientational} ordering.  

The contribution $W\Q$ to the deviatoric stress in active systems 
modifies sharply the stress vs rate flow curve. To see this,  
start with a passive sheared nematogenic system \cite{olmsted} 
in the isotropic phase near the transition to a nematic. 
Qualitatively, as the shear-rate is increased from zero, $\Q$ 
increases initially 
linearly, then more rapidly and then essentially linearly again, 
leading to shear-thinning \cite{olmstedlu}.  
If we switch on activity, with a {\em positive} value of $W$, 
the rapid increase in $\Q$ implies an equally rapid increase in $\sigac$. 
This will at the very least mitigate the shear-thinning and, if strong enough, 
will lead to shear-thickening. Alternatively, a system with $W < 0$ will enhance 
the unstable shear thinning. Note that the sign of $W$ can be got from 
an independent experiment at low concentration, simply by seeing whether 
switching on activity increases or decreases the viscosity. Thus, 
the effect of activity on the zero frequency shear viscosity predicts the 
shear-thickening or -thinning nature of the active suspension. 

We now calculate active fluctuation corrections to the shear
viscosity. 
Eq. (\ref{sigproptoq}) contributes an active force density $\sim W_2 \nabla \Q \Q$ 
to the momentum equation, whose effect on viscosities, at one-loop order, 
is of the form   
$\Delta \eta(\omega) 
\sim W \int \mbox{d}^3k \mbox{d}t \exp(i \omega t) G_{\small \Q}(k, t) C_{\small \Q}(-k, t)$,
where $G_{\small \Q}$ and $C_{\small \Q}$ are respectively the propagator 
and correlation function of $\Q$. From 
(\ref{opeom}), 
\beq
\label{deltaeta}
{\Delta \eta(\omega) \over \eta_0} \sim {W N_Q \over \eta_0 D^{3/2}(i \omega)^{1/2}} \, \, \, \,
{\mbox{for}} \,\,\, \, \omega \tau \gg 1 \, .
\eeq
Expressing the noise strength in terms of an effective 
temperature $T_{eff}$, and  
assuming on dimensional grounds $\eta_0/W \sim \tau_0$, 
$N_Q \sim k_BT_{eff}/\tau_0$, and $D \sim \ell^2/\tau_0$, 
$\tau_0 = \eta_0 \ell^3 /k_B T_{eff}$ is the rotational 
relaxation time of a single active particle, and $\ell$ its typical size, 
we see from (\ref{deltaeta}) that 
$\Delta \eta(\omega)/ \eta_0 \sim (\omega \tau_0)^{-1/2}$, 
i.e., $G^*(\omega) \sim \sqrt{i \omega}$.  

All of the above effects are likely to be greatly enhanced if the transition 
is to a polar-ordered phase, since such a transition is expected \cite{tonertu} 
to be continuous, so that $\tau$ can increase without bound. 
Furthermore, since the bare timescale $\tau_0$ is of order seconds for bacteria, 
the effects can be observed over a large dynamic range.  

Activity greatly enhances the noise temperature: on dimensional 
grounds the variance of $\sigac(k = 0, \omega = 0)$ is 
$\sim W^2 \xi^{\,3} \tau$, with $W \sim  {\eta u_0 / \xi}$ 
for active particles moving with typical speed $u_0$, correlated 
over a scale $\xi$ and time $\tau$. Equating this to   
$k_{B}T_{eff}\,\eta$ and estimating  
$\eta \sim \eta_{water} = 0.01\,$ poise, $u_0$ to be a bacterial swimming 
speed $\sim 20 \,\mu$m/s, $\,\tau \sim$ $1$ sec (an E. coli run time) gives us an 
noise temperature $T_{eff} \sim  10^5-10^6\,$K, consistent with \cite{xlwu}. 
This will mean a thousandfold enhancement of the $t^{-d/2}$ long-time tails 
\cite{tails} in 
the autocorrelation of tagged-particle velocities.  
On timescales shorter than $\tau$, effects associated with spatiotemporal 
correlations in $\Q$ \cite{gregoire} intervene. 
For a drop \cite{shiva} or a film \cite{xlwu} of size $L$ 
the tails will be cut off on a scale $\tau_v \sim \rho L^2/\pi^2\eta$.    
In \cite{xlwu}, $\tau \simeq \tau_v$. 

Rheology enters biology crucially through the active order parameter 
$\Q$ in several motility experiments which we discuss below. 
In gels imposed strains as well as elastic 
anisotropies enter (\ref{opeom}) in exactly the same way as $\A$ does 
in a fluid medium.  
This provides a natural explanation for {\it elastotaxis}, the ability of individual
motile rod-shaped bacteria such as {\it Myxococcus xanthus} to orient with their long axes along the
extension axis of an imposed elastic stress in their substrate \cite{fontes}, 
as well as {\em active mechanosensing} \cite{mechanosens}, where cells orient along 
the axis of greatest rigidity of an ambient gel.  
\begin{figure}
\includegraphics[width=7cm,height=2.5cm]{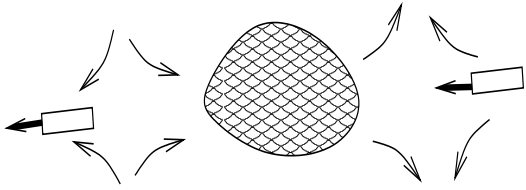}
\caption{\label{polydepoly} Flow fields due to polymerisation and depolymerisation 
at two ends of an aggregate}
\end{figure}
{\em Cytoplasmic streaming} \cite{bray}, associated with the crawling of 
amoebae, arises naturally in our model. 
Fig. \ref{polydepoly} shows that an aggregate actively polymerising at 
one end and depolymerising at the other has induced flow 
fields with extensional 
and compressional axes interchanged. 
The resulting gradient in the active stress 
can be seen, in Fig. \ref{polydepoly}, to generate a mass flux from 
left to right. 
The effect will be enhanced by the fact that the 
depolymerising end, with negative $W$, is shear thinning and hence more fluid. 
These arguments suggest why such streaming always accompanies amoeboid locomotion.  

To summarise, we have constructed the general equations governing the 
rheology of suspensions of active particles, and derived several novel 
predictions, quantitative and qualitative. 
Our description is universal:
only the values of parameters such as $W$,   
$\tau$ and $\lambda_0$ distinguish the rheologies of a 
bacterial suspension and a motor-microtubule extract. 
We look forward to tests of 
these predictions in experiments on living, reconstituted, or artificial \cite{nasseri}  
active-particle systems. 

We thank J-F. Joanny and J. Prost for interesting discussions and M. Fontes for bringing 
\cite{fontes} to our notice.
MR thanks DST, India for a Swarnajayanti grant and the Institut Curie, Paris,  
for support during an enjoyable visit, and SR acknowledges support 
from a Chaire Paris Science at ESPCI, Paris.


\begin{thebibliography}{}
\bibitem{QQ} Active particles with more elaborate complexions of forces should 
give rise to the symmetry-allowed $O(\Q^2)$ contributions in (\ref{sigproptoq}).
\end{thebibliography}
\end{document}